\documentclass[suppldata]{interact}
\usepackage{epstopdf}
\usepackage[caption=false]{subfig}
\usepackage{siunitx}
\usepackage{booktabs}
\usepackage{multirow}
\usepackage{array}
\usepackage{algpseudocode}
\usepackage{algorithm}
\usepackage{algorithmicx}
\usepackage{tabularx}
\usepackage{hyperref}
\usepackage{xcolor}
\usepackage[mathlines]{lineno}
\usepackage[version=3]{mhchem}
\usepackage[numbers,sort&compress,merge]{natbib}
\bibpunct[, ]{[}{]}{,}{n}{,}{,}

\theoremstyle{plain}
\usepackage{fix-cm}

\theoremstyle{definition}

\theoremstyle{remark}

\hypersetup{
	pdftitle={Finite-size Effects of the Excess Entropy Computed from
Integrating the Radial Distribution Function},
	pdfauthor={[Darhsan Raju]},
}
\DeclareUnicodeCharacter{0301}{\'{}}
\DeclareUnicodeCharacter{0308}{\"{}}
\begin{document}

\title{Finite-size Effects of the Excess Entropy Computed from Integrating the Radial Distribution Function}

\author{
\name{Darshan Raju\textsuperscript{a}, Mahinder Ramdin\textsuperscript{a}, Jean-Marc Simon\textsuperscript{b}, Peter Kr\"uger\textsuperscript{c}, and Thijs J.H. Vlugt\textsuperscript{a}\thanks{CONTACT Thijs J.H. Vlugt. Email: t.j.h.vlugt@tudelft.nl}}
\affil{\textsuperscript{a}Engineering Thermodynamics, Process \& Energy Department, Faculty of Mechanical Engineering, Delft University of Technology, Leeghwaterstraat 39, 2628CB Delft, The Netherlands; \textsuperscript{b}ICB, UMR 6303 CNRS, Universit\'e de Bourgogne, F-21078 Dijon, France; \textsuperscript{c}Graduate School of Engineering and Molecular Chirality Research Center, Chiba University, 1-33 Yayoi-cho, Inage, Chiba 263-8522, Japan.}}

\maketitle

\begin{abstract}
Computation of the excess entropy $S^{\rm ex}$ from the second-order density expansion of the entropy holds strictly for infinite systems in the limit of small densities. For the reliable and efficient computation of $S^{\rm ex}$ it is important to understand finite-size effects. Here, expressions to compute $S^{\rm ex}$ and Kirkwood-Buff (KB) integrals by integrating the Radial Distribution Function (RDF) in a finite volume are derived, from which $S^{\rm ex}$ and KB integrals in the thermodynamic limit are obtained. The scaling of these integrals with system size is studied. We show that the integrals of $S^{\rm ex}$ converge faster than KB integrals. We compute $S^{\rm ex}$ from Monte Carlo simulations using the Wang-Ram\'{i}rez-Dobnikar-Frenkel pair interaction potential by thermodynamic integration and by integration of the RDF. We show that $S^{\rm ex}$ computed by integrating the RDF is identical to that of $S^{\rm ex}$ computed from thermodynamic integration at low densities, provided the RDF is extrapolated to the thermodynamic limit. At higher densities, differences up to $20\%$ are observed.
\end{abstract}


\section{Introduction}
\label{sec:intro}
The excess entropy $S^{\rm ex}$ of a system of $N$ interacting molecules is defined as the difference between the entropy $S$ of the system and the entropy $S^{\rm ig}$ of an ideal gas at the same temperature $T$ and number density $\rho=N/V$, so $S^{\rm ex}(T,\rho) = S(T,\rho)-S^{\rm ig}(T,\rho)$~\cite{laird1992calculation}. The excess entropy plays a crucial role in recent theories for predicting transport properties of fluids such as diffusion coefficients, viscosities, and thermal conductivities~\cite{ghaffarizadeh2024picture,bursik2024viscosities,ghaffarizadeh2022excess, saliou2021excess,bell2019modified,hopp2019thermal,dyre2018perspective}. Hence, there is considerable interest in computing the excess entropy of systems of interacting molecules from molecular simulation. For example, this can be done by performing a free energy calculation (i.e. computing the excess free energy $A^{\rm ex}$~\cite{frenkel2023understanding}) and using the definition $A^{\rm ex} = U^{\rm ex} - TS^{\rm ex}$ in which $T$ is temperature and $U^{\rm ex}(T,\rho)=U(T,\rho)-U^{\rm ig}(T,\rho)$ is the excess potential energy of the system. A more convenient (and computationally less expensive) way is to approximate the excess entropy by a second-order density expansion of the entropy~\cite{laird1992calculation,raveche1971entropy,baranyai1989direct}. For an infinitely large system, one can derive the following approximation for the excess entropy~\cite{laird1992calculation,raveche1971entropy}

\begin{equation}
\frac{S^{\rm ex}}{k_{\rm B} N} \approx -2\pi \rho \int_0^{\infty}
\left[g(r)\ln(g(r)) - g(r) + 1 \right] r^2 {\rm d}r
\label{eq:excess}
\end{equation}

\noindent in which $k_{\rm B}$ is the Boltzmann factor and $g(r)$ is the Radial Distribution Function (RDF), which describes the local density at distance $r$ around a central molecule. As $g(r)$ is computed directly for monoatomic molecules and based on the center of mass for polyatomic molecules by most molecular simulation software, Eq.~\ref{eq:excess} provides a straightforward way to access the excess entropy of a system of $N$ interacting molecules. Eq.~\ref{eq:excess} is also used to compute excess entropies of mixtures by calculating the weighted average of the excess entropies of individual components~\citep{ghaffarizadeh2022excess,samanta2001universal,hoyt2000test}. As a result, Eq.~\ref{eq:excess} is used in screening studies~\cite{zhang2024towards,ghaffarizadeh2022excess} and crystallization studies~\cite{piaggi2017entropy,piaggi2017enhancing} to compute $S^{\rm ex}$. Including higher-order terms in the density expansion of the entropy requires 3-molecule correlation functions~\cite{baranyai1989direct,laird1992calculation} which are not often computed due to their complexity~\cite{baranyai1989direct,laird1992calculation}. In the context of liquids, the literature often highlights that the second-order density expansion of the entropy accounts for approximately 90\% of the excess entropy~\cite{piaggi2017entropy,laird1992calculation,wallace1994statistical,baranyai1989direct,wallace1987role}. Recently, Huang and Widom~\cite{huang2024entropy} computed entropies from the third-order density expansion of the entropy using 3-molecule correlation functions following the Kirkwood and Boggs superposition approximation~\cite{kirkwood1942radial}. These authors show that the third-order density expansion of entropy marginally enhance the accuracies in estimating $S^{\rm ex}$ compared to the second-order expansion.

\medskip

In this paper, we investigate in detail the underlying approximations of Eq.~\ref{eq:excess} to compute the excess entropy: {\em (1)} Eq.~\ref{eq:excess} is a low-density approximation~\cite{baranyai1989direct} so at high densities one would expect deviations from the exact value of $S^{\rm ex}$; {\em (2)} RDFs computed by molecular simulations shows finite-size effects~\cite{kruger2018size,ganguly2013convergence,salacuse1996finite}, e.g. $g(r)$ approaches $1$ at large distances $r$ only if very large systems are considered. This may influence the computed value of $S^{\rm ex}$; {\em (3)} Similar to Kirkwood-Buff (KB) integrals~\cite{kruger2018size,kruger2013kirkwood}, Eq.~\ref{eq:excess} is valid only for infinite systems and it is not a priori clear if it is allowed to truncate the integration of Eq.~\ref{eq:excess} at finite distances. In section~\ref{sec:KB}, we investigate the truncation of Eq.~\ref{eq:excess} using an analytic model function for $g(r)$, and we will show that this truncation is possible provided that the range of $g(r)$ is not too long. In the next sections, we systematically investigate the other two assumptions by comparing them with molecular simulations. Simulation details are provided in section~\ref{sec:sims}, and a detailed analysis of finite-size effects is provided in section~\ref{sec:results}. Our main findings are summarized in section~\ref{sec:conclusions}.

\section{Truncation of the integral for \texorpdfstring{$S^{\rm ex}$}{Sex}}
\label{sec:KB}

It is important to note that Eq.~\ref{eq:excess} is strictly speaking only valid for infinite systems. For a finite system with a volume $V$, to obtain $S^{\rm ex}$ one has to integrate the function $\left[g(r)\ln(g(r)) - g(r) + 1\right]$ over the positions of two particles ${\bf r}_1$ and ${\bf r}_2$ inside this volume $V$~\cite{kruger2013kirkwood}. Only when the volume $V$ is infinitely large, one can replace the integral over the positions ${\bf r}_1$ and ${\bf r}_2$ by an integral over their distance $r=\left| {\bf r}_1-{\bf r}_2\right|$ between ${\bf r}_1$ and ${\bf r}_2$. For $S^{\rm ex}$, we have for a spherical volume $V$ with diameter $L$~\cite{kruger2013kirkwood}

\begin{equation}
\begin{aligned}
  \frac{S^{\rm ex}}{k_{\rm B} N} &  \approx -\frac{\rho}{2}\int_V\int_V
  \left[g(r)\ln(g(r)) - g(r) + 1 \right] {\rm d}{\bf r}_1 {\rm d}{\bf
    r}_2 \\ & = -\frac{\rho}{2}\int_0^L
  w(r,L) \left[g(r)\ln(g(r)) - g(r) + 1 \right] {\rm d}r
  \label{eq:excess2}
\end{aligned}
\end{equation}
in which
\begin{equation}
\begin{aligned}
w(r,L) & = 4\pi r^2 \left[1-\frac{3}{2}\left(\frac{r}{L}\right) +
  \frac{1}{2}\left(\frac{r}{L}\right)^3 \right]
\end{aligned}
\end{equation}

\noindent is the geometric weight function for a sphere with diameter $L$~\cite{kruger2018size}. Clearly, in the limit, $L\rightarrow\infty$, Eq.~\ref{eq:excess2} reduces to Eq.~\ref{eq:excess}. For finite-size systems, this is not valid, so one must strictly use Eq.~\ref{eq:excess2} instead of Eq.~\ref{eq:excess}. This finite-size effect was derived first in the context of Kirkwood-Buff (KB) integrals~\cite{kruger2018size,kruger2013kirkwood} where one has to integrate $\left[g(r)-1\right]$ over the positions of two particles ${\bf r}_1$ and ${\bf r}_2$ inside volume $V$, rather than integrating the function $\left[g(r)\ln(g(r)) - g(r) + 1\right]$. For simplicity, let us define the integral over positions ${\bf r}_1$ and ${\bf r}_2$ in volume $V$

\begin{equation}
X(L) = \int_V \int_V q(r)  {\rm d}{\bf r}_1 {\rm d}{\bf r}_2 =
4\pi\int_0^Lq(r)\left[ 1-\frac{3}{2}\left(\frac{r}{L}\right) +
  \frac{1}{2}\left(\frac{r}{L}\right)^3\right]r^2{\rm d}r.
\end{equation}
We also define
\begin{equation}
X^*(L) = 4\pi\int_0^Lq(r)r^2{\rm d}r,
\label{eq:5}
\end{equation}

\noindent which is commonly referred to as the running integral. Only in the limit $L\rightarrow\infty$, $X(L)$ can be replaced by $X^*(L)$. We are interested in an estimation of the value $X$ in the thermodynamic limit (i.e. $L\rightarrow\infty$) which we will denote by $X_\infty$, obtained by extrapolating from a system at finite volume $V$. In case of KB coefficients, we have $q_{\rm KB}(r)=\left[g(r)-1\right]$ and for the excess entropy we have $q_{S}(r)=\left[g(r)\ln(g(r))-g(r)+1\right]$. It is important to note that for large distances $r$, the scaling behavior of these properties is different. As for large distances $r$, $g(r)$ is close to $1$, we can write $g(r)=1+\delta$, where $|\delta|<<1$, leading to $q_{\rm KB}\approx\delta$ and $q_{S}\approx \delta^2$. This indicates that the convergence of KB integrals will generally be much more difficult than the integrals for computing the excess entropy.

\medskip

In Ref.~\cite{kruger2013kirkwood} it was shown that for a finite-correlation length of $q(r)$, the value of $X_\infty$ can be approximated by a Taylor expansion in $1/L$ and only the first-order derivative was considered. We can write the approximation up to the third-order as
\begin{equation}
  X_\infty = X(L) - \frac{1}{L}\frac{{\rm d}X(L)}{{\rm d}(1/L)} +
  \frac{1}{2L^2}\frac{{\rm d}^2X(L)}{{\rm d}(1/L)^2} -
  \frac{1}{6L^3}\frac{{\rm d}^3X(L)}{{\rm d}(1/L)^3} + \mathcal{O}(1/L^4).
  \label{eq:taylorexpansion}
\end{equation}
Using the Leibniz rule~\cite{amazigo1980advanced}, we find for the derivatives
\begin{align}
  \frac{{\rm d}X(L)}{{\rm d}(1/L)}& =4\pi\int_0^L\left[ -\frac{3r}{2} + \frac{3r^3}{2L^2}\right]q(r)r^2{\rm d}r,
\\[1em]
  \frac{{\rm d}^2X(L)}{{\rm d}(1/L)^2} & = 4\pi\int_0^L \frac{3r^3}{L}q(r)r^2{\rm d}r,
\\[1em]
  \frac{{\rm d}^3X(L)}{{\rm d}(1/L)^3} & = -12\pi L^6q(L) + 4\pi\int_0^L 3r^3q(r)r^2{\rm d}r.
\end{align}
By substitution of these expressions into Eq.~\ref{eq:taylorexpansion}, we obtain approximations for $X_\infty$ of different order
\begin{align}
X_\infty^1 & = X(L) - \frac{1}{L}\frac{{\rm d}X(L)}{{\rm d}(1/L)} \notag \\ 
& =4\pi\int_0^L\left[1-\frac{r^3}{L^3} \right]q(r)r^2{\rm d}r,
\label{eq:10}
\\[1em]
X_\infty^2 & = X(L) - \frac{1}{L}\frac{{\rm d}X(L)}{{\rm d}(1/L)} +
\frac{1}{2L^2}\frac{{\rm d}^2X(L)}{{\rm d}(1/L)^2} \notag \\ 
& =4\pi\int_0^L\left[1+\frac{r^3}{2L^3} \right]q(r)r^2{\rm d}r,
\\[1em]
X_\infty^3  & = X(L) - \frac{1}{L}\frac{{\rm d}X(L)}{{\rm d}(1/L)} +
\frac{1}{2L^2}\frac{{\rm d}^2X(L)}{{\rm d}(1/L)^2}-\frac{1}{6L^3}\frac{{\rm d}^3X(L)}{{\rm d}(1/L)^3} \notag \\ 
& =4\pi\int_0^L q(r)r^2{\rm d}r + 2\pi L^3q(L) = X^*(L) +  2\pi L^3q(L).
\end{align}

\noindent We are not considering even higher-order derivatives, as these would involve derivatives of $q(L)$ with respect to $L$. The third-order approximation $X^3_{\infty}$ includes the running integral shown in Eq.~\ref{eq:5}, which is often used at large $L$, as an alternative to the KB and excess entropy integrals in the thermodynamic limit~\cite{ben2006molecular}. For KB integrals, it was previously found that the first-order approximation $X_\infty^1$ provides accurate results and that higher-order derivatives can be neglected~\cite{kruger2013kirkwood}, clearly showing that $X(L)$ scales nearly linearly with $1/L$. 

\medskip

\begin{figure}[tbh!]
    \captionsetup{justification=raggedright,singlelinecheck=false}
    \centering
     \captionsetup[subfigure]{position=top,labelfont={normal},textfont=normal,singlelinecheck=off,justification=raggedright}
    \subfloat[]{\includegraphics[width=12cm,height=6cm]{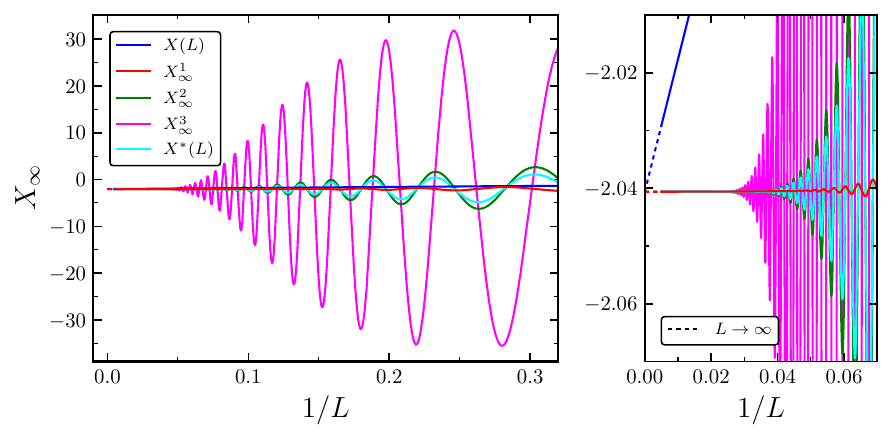}}
    \label{fig1a}
    \subfloat[]{\includegraphics[width=12cm,height=6cm]{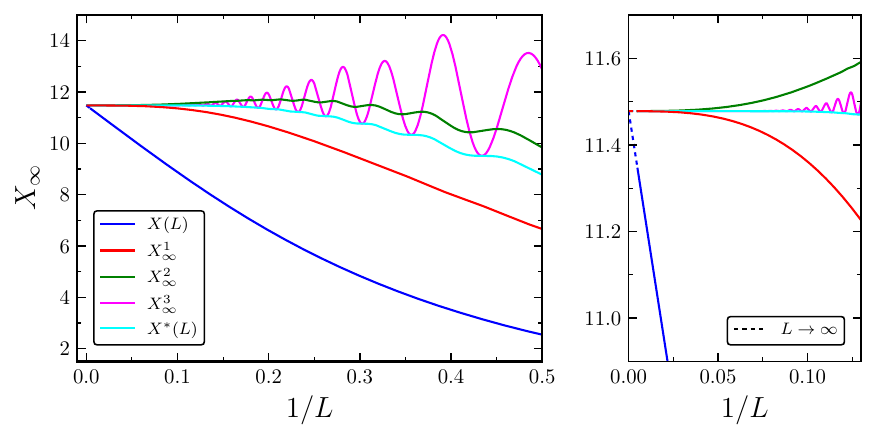}}
    \label{fig1b}
    \caption{Comparison of different approximations $X(L)$, $X^1_{\infty}$, $X^2_{\infty}$, $X^3_{\infty}$, and  $X^*(L)$ to compute (a) Kirkwood-Buff (KB)  integrals and (b) excess entropy ($S^{\mathrm{ex}}$) integrals in the thermodynamic limit ($L\rightarrow\infty$) obtained from the analytic Radial Distribution Function (RDF)~\cite{kirkwood1942radial,verlet1968computer} for $\chi=2$.}
    \label{fig1}
\end{figure}

\begin{figure}[tbh!]
    \captionsetup{justification=raggedright,singlelinecheck=false}
    \centering
     \captionsetup[subfigure]{position=top,labelfont={normal},textfont=normal,singlelinecheck=off,justification=raggedright}
    \subfloat[]{\includegraphics[width=12cm,height=6cm]{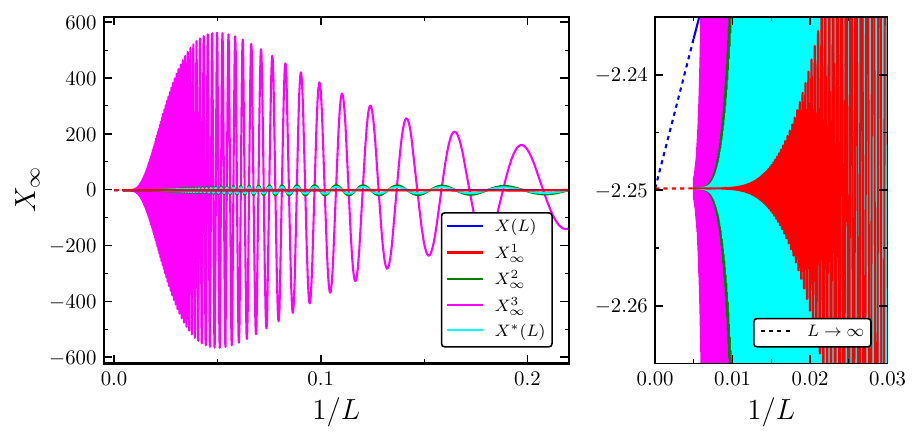}}
    \label{fig2a}
    \subfloat[]{\includegraphics[width=12cm,height=6cm]{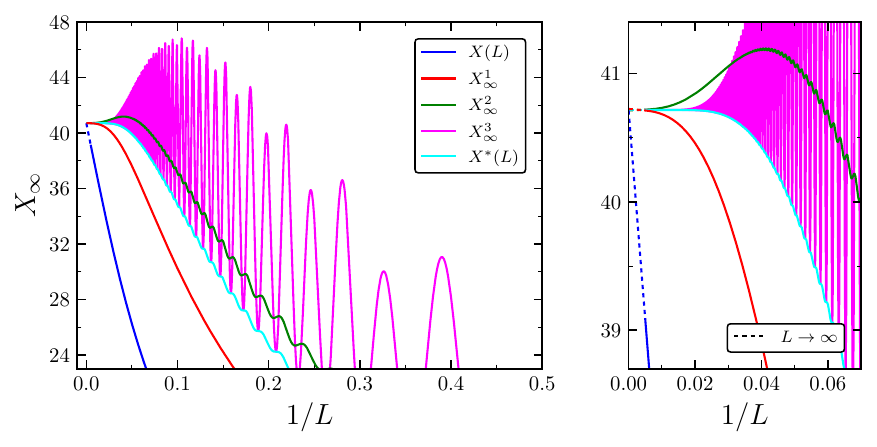}}
    \label{fig2b}
    \caption{Comparison of different approximations $X(L)$, $X^1_{\infty}$, $X^2_{\infty}$, $X^3_{\infty}$, and  $X^*(L)$ to compute (a) Kirkwood-Buff (KB)  integrals and (b) excess entropy ($S^{\mathrm{ex}}$) integrals in the thermodynamic limit ($L\rightarrow\infty$) obtained from the analytic Radial Distribution Function (RDF)~\cite{kirkwood1942radial,verlet1968computer} for $\chi=10$.}
    \label{fig2}
\end{figure}

\begin{figure}[tbh!]
    \captionsetup{justification=raggedright,singlelinecheck=false}
    \centering
     \captionsetup[subfigure]{position=top,labelfont={normal},textfont=normal,singlelinecheck=off,justification=raggedright}
    \subfloat[]{\includegraphics[width=12cm,height=6cm]{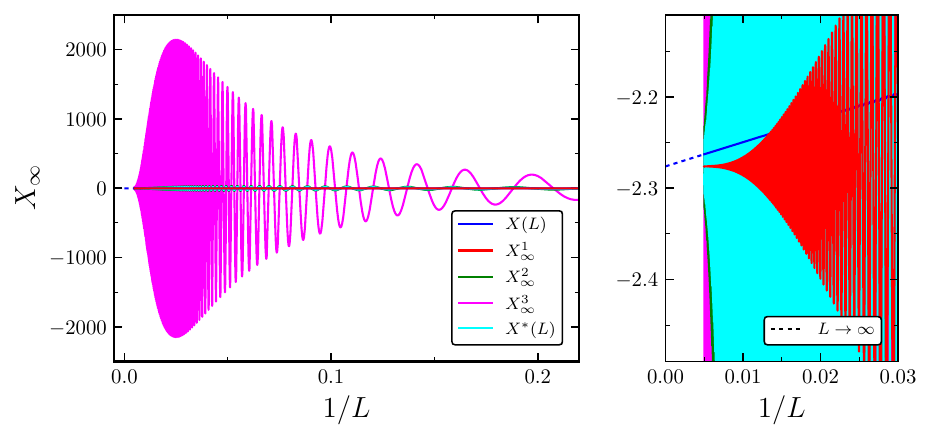}}
    \label{fig3a}
    \subfloat[]{\includegraphics[width=12cm,height=6cm]{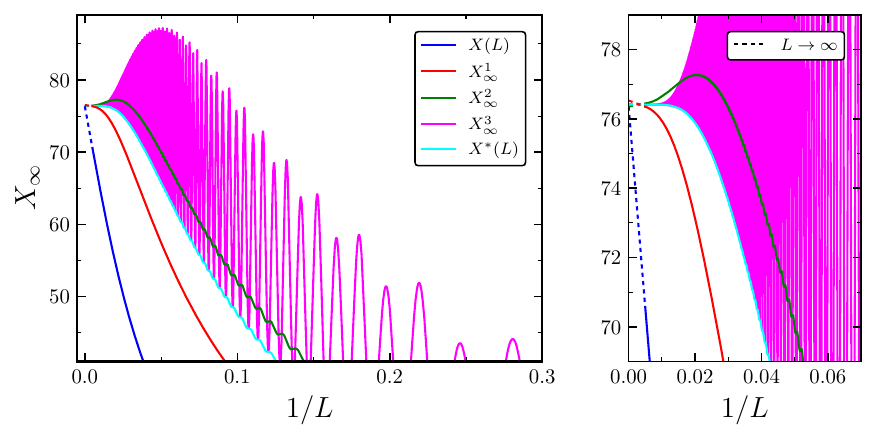}}
    \label{fig3b}
    \caption{Comparison of different approximations $X(L)$, $X^1_{\infty}$, $X^2_{\infty}$, $X^3_{\infty}$, and  $X^*(L)$ to compute (a) Kirkwood-Buff (KB)  integrals and (b) excess entropy ($S^{\mathrm{ex}}$) integrals in the thermodynamic limit ($L\rightarrow\infty$) obtained from the analytic Radial Distribution Function (RDF)~\cite{kirkwood1942radial,verlet1968computer} for $\chi=20$.}
    \label{fig3}
\end{figure}

To test the various estimates for $X_\infty$ (for both KB ($q_{\rm KB}(r)$) and excess entropy ($q_{S}(r)$) integrals), i.e., the value of $X$ in the thermodynamics limit, we consider an analytic model for the RDF: $g(r) = 1 + 3/2\exp[(1-r)/\chi]\cos(2\pi(r-(21/20)))/r$ for $r> (19/20)$ and $g(r)=0$ otherwise~\cite{kirkwood1942radial,verlet1968computer,kruger2018size}. The parameter $\chi$ controls the range of the interactions. This approach allows one to separately consider finite-size effects of the integral and the other finite-size effects. Fig.~\ref{fig1} shows different orders of approximation of $X_{\infty}$, along with exact ($X(L)$) and running integral ($X^*(L)$), for $q_{\rm KB}(r)$ and $q_{S}(r)$ using the analytic expression of $g(r)$ for $\chi = 2$. As shown in Fig.~\ref{fig1}a for KB integrals, the exact expression, ($X(L)$), is free from oscillations but achieves convergence only at a very large length scale compared to $X^*(L)$ and different order of approximation. The third-order approximation of the KB integral has large oscillations and poor convergence compared to other approximations, as shown in Fig.~\ref{fig1}a. The amplitude of these oscillations decreases with lower approximation orders, while the running KB integral ($X^*(L)$) has an oscillation amplitude slightly less than the second-order approximation. $X^3_{\infty}$ reaches an asymptotic value at $L$ $\approx$ $40$, while $X^2_{\infty}$ and $X^*(L)$ converges at $L$ $\approx$ $30$. The first-order approximation has the lowest amplitude of oscillations and converges to an asymptotic value at  $L$ $\approx$ $20$. The observation that the first-order approximation converges better than $X(L)$ and $X^*(L)$ aligns with the work of Kr{\"u}ger et al.~\cite{kruger2013kirkwood} for KB integrals. For excess entropy integrals shown in Fig.~\ref{fig1}b, it is clear that the $X(L)$ suffers from poor convergence compared to different order of approximations and $X^*(L)$. Unlike KB integrals, the first and second-order approximations show no oscillations, while the third-order approximation does suffer from oscillations. The running excess entropy integral, $X^*(L)$, has minor oscillations, with its function intersecting the local minima of $X^3_{\infty}$. $X^1_{\infty}$ of the excess entropy integral reaches an asymptotic value at $L$ $\approx$ $50$, while $X^2_{\infty}$ reaches at $L$ $\approx$ $40$. $X^*(L)$ and $X^3_{\infty}$ approaches an asymptotic value at $L$ $\approx$ $10$, but $X^3_{\infty}$ has minor oscillations for $L$ $<$ $10$. Convergence of the excess entropy integral increases with increasing order of approximations, and $X^*(L)$ has better convergence compared to different order of approximations ($X^1_{\infty}$, $X^2_{\infty}$, $X^3_{\infty}$). The convergence of different approximations for KB integrals shown in Fig.~\ref{fig2}a and Fig.~\ref{fig3}a for $\chi$ $=$ $10$ and $\chi$ $=$ $20$ respectively follows the order $X^1_{\infty}$ $>$ $X^*(L)$ $>$ $X^2_{\infty}$ $>$ $X^3_{\infty}$, similar to $\chi$ $=$ $2$. $X^1_{\infty}$ converges to an asymptotic value at $L$ $\approx$ $130$ and $L$ $>$ $200$ for $\chi$ $=$ $10$ and $\chi$ $=$ $20$ respectively. Similarly, for the excess entropy, the convergence is achieved at a smaller length scale $L$ in the following order $X^*(L)$ $>$ $X^3_{\infty}$ $>$ $X^2_{\infty}$ $>$ $X^1_{\infty}$ for both $\chi$ $=$ $10$ and $20$ as shown in Fig.~\ref{fig2}b and Fig.~\ref{fig3}b. $X^*(L)$ converges to an asymptotic value at $L$ $\approx$ $50$ and $L$ $\approx$ $100$ for $\chi$ $=$ $10$ and $\chi$ $=$ $20$ respectively. The foregoing comparison of the minimum $L$ value needed for integral convergence shows that the minimum $L$ value needed for the KB integral ($q_{\rm KB}$ with $X^1_{\infty}$) is always at least twice the minimum $L$ value needed for the excess entropy ($q_{S}$ with $X^*(L)$).

\medskip

\begin{figure}
    \centering
    \includegraphics[width=0.75\linewidth]{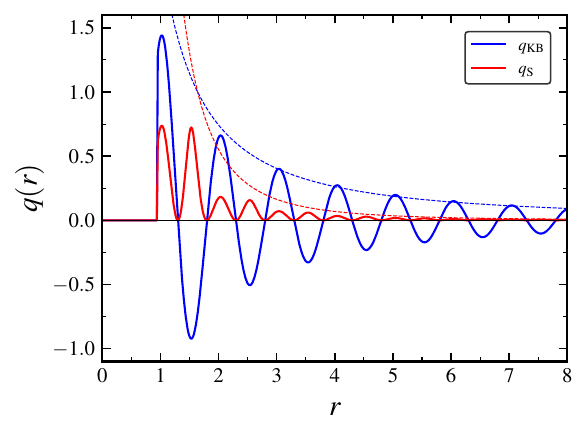}
    \caption{Functions $q_{\mathrm{KB}}(r)$ (thick blue line) and $q_{S}(r)$ (thick red line) for the analytic Radial Distribution Function (RDF) with $\chi = 10$. The dotted lines are guides to the eye to indicate the amplitude decrease. The blue dotted line is $2.1/r^{3/2}$ (fit line to the maxima) and the red dotted line is the square of the blue one.}
    \label{fig4}
\end{figure}

We have observed that for the KB integrals, i.e., the function $q_{\rm KB}(r)$, the first order extrapolation $X_\infty^1$ has the best convergence properties; in particular, it improves with respect to $X^*(L)$. This finding has been discussed before~\cite{kruger2013kirkwood,dawass2018kirkwood} and can be understood from the fact that the weight function $4\pi r^2$ in $X^*(L)$ considerably amplifies the sign-changing oscillations of $q_{\rm KB}(r)$ (see Fig.~\ref{fig4}, blue line)~\cite{santos2018finite}. Therefore, simple truncation of the integral at~$L$ gives rise to large oscillations of $X^*(L)$ and thus slow convergence (Fig.~\ref{fig1}a). In contrast, $X_\infty^1$ is based on the exact and almost oscillation-free finite volume KB integral $X(L)$. The difference $X(L)-X_\infty$, which may be considered as a surface term~\cite{dawass2020kirkwood}, is known to scale as $1/L$ for large~$L$~\cite{kruger2013kirkwood,schnell2011calculating}. In $X_\infty^1$, the leading error of $X(L)$ is corrected without much deteriorating the smoothness inherited from $X(L)$, because the weight function $4\pi r^2[1-(r/L)^3]$ (see Eq.~\ref{eq:10}) is continuous at $r=L$~\cite{kruger2018size}. From the present analysis, it is seen that the extrapolations based on higher order Taylor expansion $X_\infty^2$ and $X_\infty^3$ lead to a quite strong amplification of oscillations in $q_{\rm KB}(r)$, and thus to a slower convergence than $X_\infty^1$ and $X^*(L)$. For the excess entropy, however, we see that the fastest convergence is obtained with the running integral $X^*(L)$ rather than $X_\infty^1$. So the question arises: why does the above reasoning, which explains the convergence behavior of integrals over $q_{\rm KB}(r)$, not hold when the integrand is $q_S(r)$? As seen from Fig.~\ref{fig4}, the function $q_S(r)$ oscillates, but with a much weaker amplitude than $q_{\rm KB}(r)$. $q_S(r)$ is an essentially positive function, while $q_{\rm KB}(r)$ changes sign at each oscillation. Both differences can easily be understood in the limit of sufficiently large $r$ (typically $r>2$), where $|g(r)-1|<1$. We define $h(r)=g(r)-1$ and have $q_S = (1+h)\ln(1+h)-h\approx h^2+{\cal O}(h^3)$. So $q_{S}(r)$ is essentially positive and, for $r\rightarrow\infty$, where $h(r)\rightarrow 0$, its amplitude is much smaller than that of $q_{\rm KB}(r)\equiv h(r)$. Since $0 < q_S(r) < 1$ and $0 < 1-(r/L)^3 <1$, it follows that the integrand of $X_\infty^1$ is smaller than that of $X^*(L)$ for all $r$. As a consequence both $X_\infty^1$ and $X^*(L)$ are strictly increasing and we have $0 < X_1(L) < X^*(L) < X_\infty$. This proves that $X^*(L)$ converges faster to $X_\infty$ than $X_\infty^1$, as also seen in the numerical result of Fig.\ref{fig1}b.

\section{Simulation details}
\label{sec:sims}

\begin{figure}[tbh!]
    \captionsetup{justification=raggedright,singlelinecheck=false}
    \centering
     \captionsetup[subfigure]{position=top,labelfont={normal},textfont=normal,singlelinecheck=off,justification=raggedright}
    \subfloat[]{\includegraphics[width=12cm]{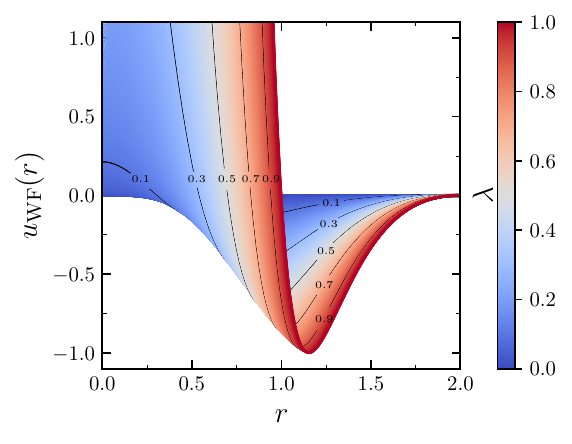}}
    \label{fig5a}
    \subfloat[]{\includegraphics[width=12cm]{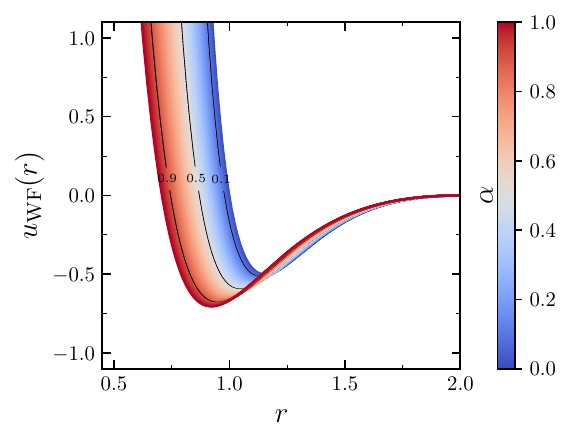}}
    \label{fig5b}
    \caption{The potential energy function $u_{\mathrm{WF}}(r,\lambda)$ as a function of distance $r$ for the Wang-Ram\'{i}rez-Dobnikar-Frenkel (WF) pair potential (Eq.~\ref{eq:15})~\cite{wang2020lennard}, with $r_c$ $=$ $2$. (a) Contour plot of WF potential for $\lambda$ ranging from $0$ to $1$ with $\alpha$ $=$ $1$. (b) Contour plot of WF potential for $\alpha$ ranging from $0$ to $1$ when $\lambda$ $=$ $0.5$.}
    \label{fig5}
\end{figure}

We consider a system of molecules in the $NVT$ ensemble that interact via the Wang-Ram\'{i}rez-Dobnikar-Frenkel (WF) pair interaction potential~\cite{wang2020lennard}
\begin{equation}
u_{\mathrm{WF}}(r) = \epsilon \left[\left(\frac{\sigma}{r}
  \right)^2-1\right]\left[\left(\frac{r_{\rm c}}{r} \right)^2
  -1\right]^2 \quad \quad r<r_{\rm c}
\end{equation}
and $u_{\mathrm{WF}}(r)=0$ otherwise. In this equation, $r$ is the distance between two interacting molecules, $\sigma$ is the size parameter, $\epsilon$ is a measure of the well-depth of the potential energy, and $r_{\rm c}$ is the cut-off radius. In the remainder of this manuscript, we will use $\sigma$ as the unit of length and $\epsilon$ as a unit of energy, so we have
\begin{equation}
u_{\mathrm{WF}}(r) =  \left[\left(\frac{1}{r}
  \right)^2-1\right]\left[\left(\frac{r_{\rm c}}{r} \right)^2
  -1\right]^2 \quad \quad r<r_{\rm c}.
\label{eq:wf}
\end{equation}
For $r_{\rm c}=2$, the WF pair potential is Lennard-Jones-like, while for $r_{\rm c}=1.2$, it behaves like typical short-range interactions between colloids~\cite{wang2020lennard,frenkel2023understanding}. The advantage of this interaction potential (e.g., compared to Lennard-Jones) is that one does not need truncation or tail corrections~\cite{frenkel2023understanding}. To compute the excess free energy of the WF system, we adopt a soft-core version of Eq.~\ref{eq:wf}
\begin{equation}
  u_{\mathrm{WF}}(r,\lambda) = \lambda \left[\frac{1}{r^2 + \alpha(1-\lambda)}
  - 1 \right]\left[\frac{r_{\rm c}^2 + \alpha(1-\lambda)}{r^2 +
      \alpha(1-\lambda)} -1\right]^2 \quad \quad r<r_{\rm c},
      \label{eq:15}
\end{equation}
where $\lambda$ is the scaling parameter that scales the strength of the WF potential. It is easy to see that for $\lambda=0$, we have an ideal gas ($u_{\mathrm{WF}}(r,\lambda=0)=0$), while for $\lambda=1$, the original WF interaction potential is recovered. For all $\lambda$, we have $u_{\mathrm{WF}}(r_{\rm c},\lambda)=0$. The parameter $\alpha$ is chosen such that one does not have a singularity for $r\rightarrow 0$ unless $\lambda=1$. Fig.~\ref{fig5}a shows the soft-core WF interaction potential, $u_{\mathrm{WF}}(r)$, plotted as a function of $r$, with  $r_c =2$ and $\alpha = 1$ for $\lambda \in [0, 1] $. For decreasing $\lambda$, the repulsive interactions become less steep and increase the interaction range over a broad distance, $r$. Similarly, as shown in Fig.~\ref{fig5}b for varying $\alpha$ with $r_c =2$ and $\lambda = 0.5$, decreasing the value of $\alpha$ also reduces the steepness of the repulsive interactions. The effective interaction range and strength can modify the conditions for a possible vapor-liquid phase transition~\cite{hens2020brick,polat2021new}. The excess free energy of a system is computed by Thermodynamic Integration (TI) by scaling the interactions of all particle pairs in the system~\cite{frenkel2023understanding}
\begin{equation}
  A^{\rm ex} = A(T,\rho) - A^{\rm ig}(T,\rho) = \int_0^1 \left\langle \left(\frac{\partial
        U(r,\lambda)}{\partial\lambda} \right) \right\rangle {\rm d}\lambda
  \label{eq:thermointegral}
\end{equation}
with $U_{\mathrm{WF}}(r,\lambda) = \sum_{i<j} u_{\mathrm{WF}}(r,\lambda)$ and
\begin{multline}
\frac{\partial u_{\mathrm{WF}}(r,\lambda)}{\partial \lambda} =
\frac{\left(r_{\rm c}^2 - r^2\right)\left[\alpha\lambda\left(r_{\rm c}^2-r^2 \right) -2\alpha\lambda\left(r_{\rm c}^2-r^2\right)\left(\alpha(1-\lambda)+r^2 - 1\right)\right]}{\left(\alpha(1-\lambda) + r^2\right)^4} \times 
\\
\frac{\left(r_{\rm c}^2 - r^2\right)\left[-\left(r_{\rm c}^2-r^2\right)\left(\alpha(1-\lambda) + r^2\right)\left(\alpha(1-\lambda)+r^2-1\right)\right]}{\left(\alpha(1-\lambda) + r^2\right)^4}.
\end{multline}

\noindent For $\lambda=1$ we can write $U^{\rm ex} = U$ and $A^{\rm ex} = U^{\rm ex}-TS^{\rm ex}$ so the excess entropy follows directly from this.

\medskip

All $NVT$ simulations were performed using an in-house Monte Carlo code. Monte Carlo trial moves consist of (randomly selected) particle displacements. Typically, $10^3$ equilibration cycles (starting from a random initial configuration) were used, and $10^6$ production cycles, with $N$ trial moves per cycle. The maximum particle displacement was adjusted to have ca. $50\%$ of all displacements accepted and was maximized to half the box size. Thermodynamic integration of Eq.~\ref{eq:thermointegral} was performed by running $100$ simulations between $\lambda=0$ and $\lambda=1$ and by fitting a spline function to $\left\langle \partial U_{\mathrm{WF}}/\partial\lambda\right\rangle $ as a function of $\lambda$. It was carefully checked that the TI does not cross any vapor-liquid phase transition. For density $\rho$, 10 independent simulations with a different initial configuration are performed to compute $S^{\rm ex}$ using Eq.~\ref{eq:excess} truncated to a finite-size. These 10 simulations are divided into 5 blocks from which average values and uncertainties of $S^{\rm ex}$ are computed. The mean and standard deviation of 5 blocks is the average value and uncertainty of $S^{\rm ex}$. Finite-size effects of $g(r)$ are corrected by the method of Ganguly and van der Vegt~\cite{ganguly2013convergence}
\begin{equation}
  g^\infty(r) = g(r) \times \frac{N  \left(1 - \frac{\frac{4}{3}\pi
      r^3}{V}\right)}{N\left( 1-\frac{\frac{4}{3}\pi r^3}{V}\right)-
    \frac{4\pi N}{V}\int_{0}^r \left[g(r')-1 \right]r^2 {\rm d}r' - 1}
    \label{eq:18}
\end{equation}
in which $g(r)$ is the RDF from a simulation of a finite system in the $NVT$ ensemble, and $g^\infty(r)$ is its estimate in the thermodynamic limit. Essentially, this method corrects for the slightly different density outside a sphere with radius $r$ around a central particle, compared to the average density $N/V$. Ganguly and van der Vegt~\cite{ganguly2013convergence} showed that the finite-size correction of RDF to the thermodynamic limit (Eq.~\ref{eq:18}) is effective for non-ideal systems with a limited number of particles. One can show that this method corrects the RDF of an ideal gas ($g(r)=(N-1)/N$) to the result in the thermodynamic limit ($g(r)=1$) and also provides a good estimation of $g(r)$ in the thermodynamic limit for non-ideal systems.

\section{Results and Discussion}
\label{sec:results}

\begin{figure}[tbh!]
    \captionsetup{justification=raggedright,singlelinecheck=false}
    \centering
     \captionsetup[subfigure]{position=top,labelfont={normal},textfont=normal,singlelinecheck=off,justification=raggedright}
    \subfloat[]{\includegraphics[width=7cm]{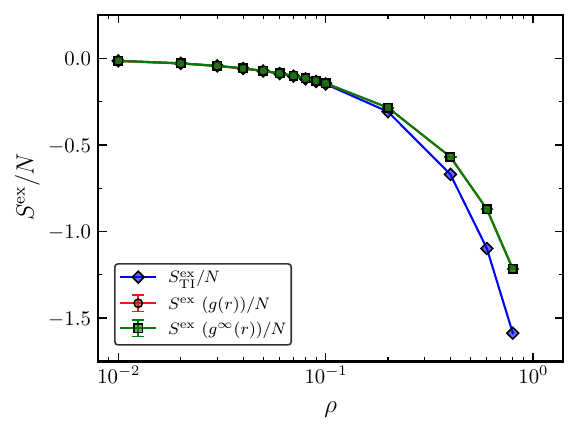}}
    \label{fig6a}
    \subfloat[]{\includegraphics[width=7cm]{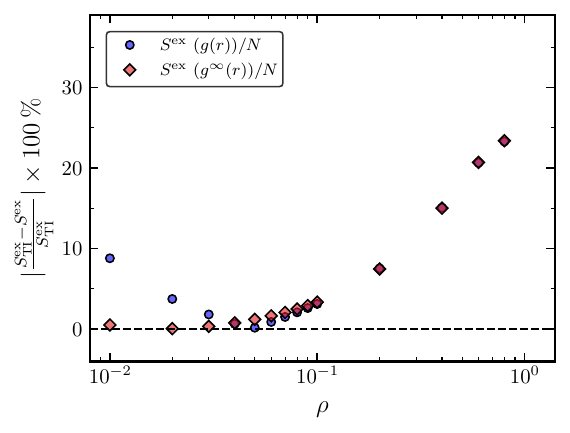}}
    \label{fig6b}
    \subfloat[]{\includegraphics[width=7cm]{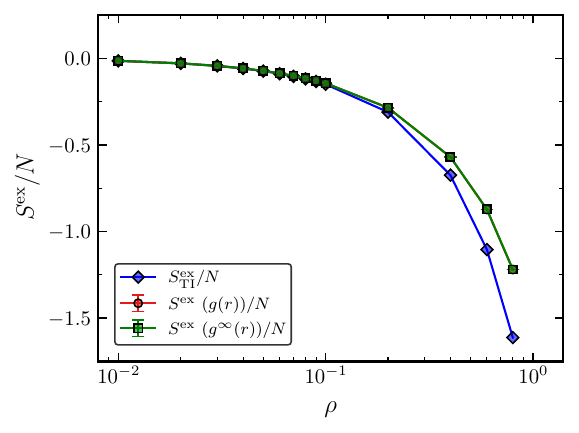}}
    \label{fig6c}
    \subfloat[]{\includegraphics[width=7cm]{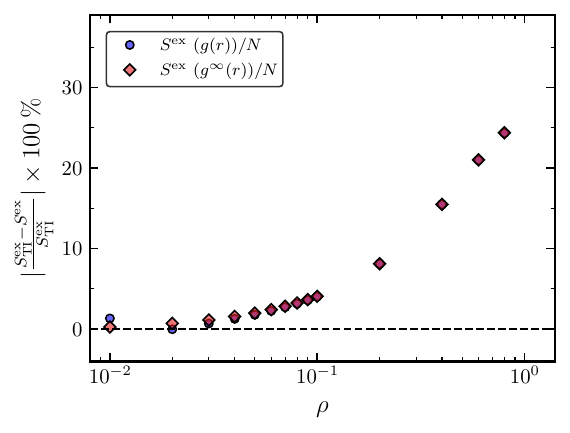}}
    \label{fig6d}
    \caption{Comparison of excess entropies $S^{\mathrm{ex}}$ computed for various densities $\rho$ using Thermodynamic Integration (TI) and Eq.~\ref{eq:excess}, with and without finite-size corrections to the RDF, $g^{\infty}(r)$ and $g(r)$, in the $NVT$ ensemble for $T = 4$, $r_c = 2$, $\alpha = 1$, and for different system size $N$: (a) Comparison of $S^{\mathrm{ex}}$ for $N = 100$. (b) Computed Absolute Percentage Error (APE) of $S^{\mathrm{ex}}$ for $N = 100$. (c) Comparison of $S^{\mathrm{ex}}$ for $N = 500$. (d) Computed APE of $S^{\mathrm{ex}}$ for $N = 500$.}
    \label{fig6}
\end{figure}

\begin{figure}[tbh!]
    \captionsetup{justification=raggedright,singlelinecheck=false}
    \centering
     \captionsetup[subfigure]{position=top,labelfont={normal},textfont=normal,singlelinecheck=off,justification=raggedright}
    \subfloat[]{\includegraphics[width=7cm]{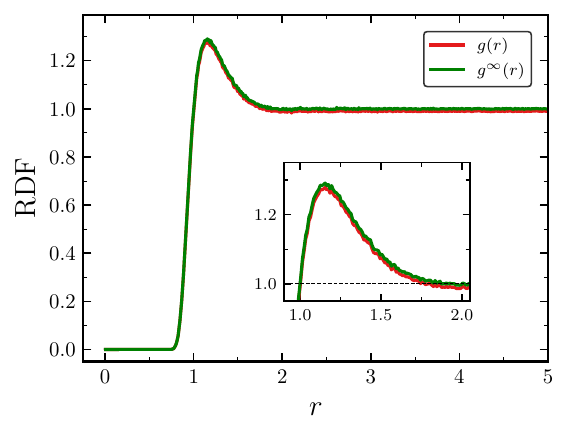}}
    \label{fig-RPa}
    \subfloat[]{\includegraphics[width=7cm]{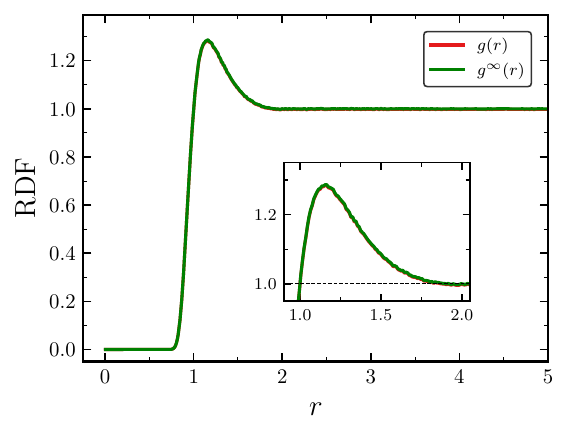}}
    \label{fig-RPb}
    \caption{Comparison of Radial Distribution Functions (RDFs) computed in the $NVT$ ensemble for $T = 4$, $r_c = 2$, $\alpha = 1$, and $\rho = 0.01$ without finite-size corrections ($g(r)$) and with finite-size corrections ($g^{\infty}(r)$) for different system size $N$. (a) Comparison of $g(r)$ and $g^{\infty}(r)$ for $N = 100$. (b) Comparison of  $g(r)$ and $g^{\infty}(r)$ for $N = 500$.}
    \label{fig-RevisionPlot}
\end{figure}

\begin{figure}[tbh!]
    \captionsetup{justification=raggedright,singlelinecheck=false}
    \centering
     \captionsetup[subfigure]{position=top,labelfont={normal},textfont=normal,singlelinecheck=off,justification=raggedright}
    \subfloat[]{\includegraphics[width=7cm]{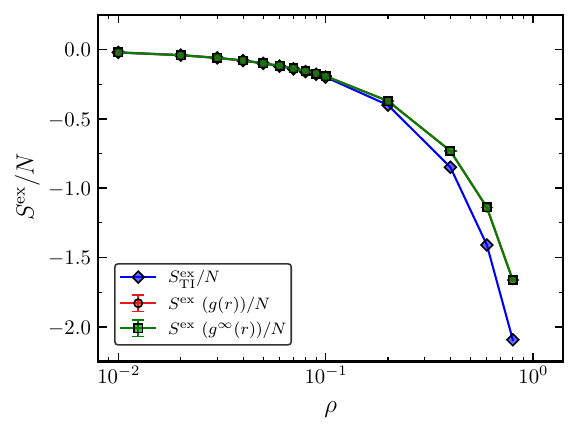}}
    \label{fig7a}
    \subfloat[]{\includegraphics[width=7cm]{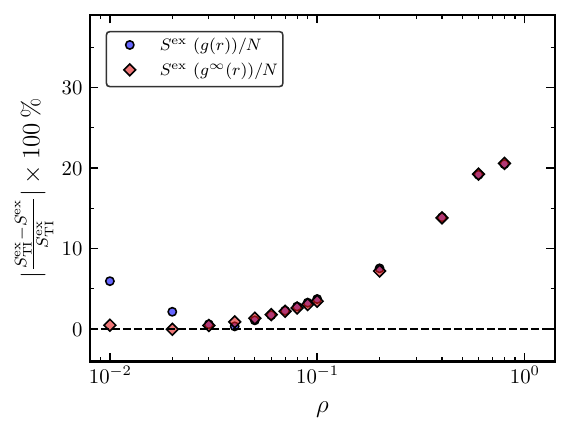}}
    \label{fig7b}
    \subfloat[]{\includegraphics[width=7cm]{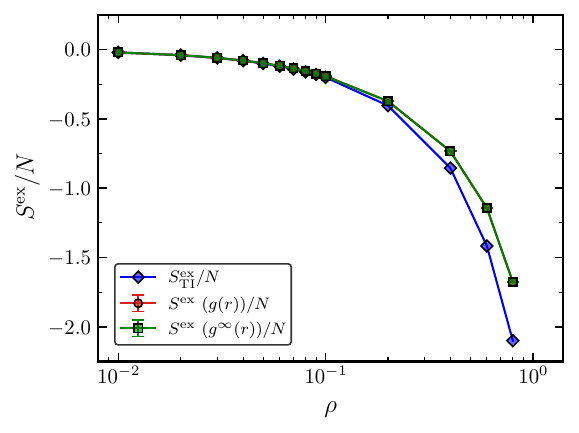}}
    \label{fig7c}
    \subfloat[]{\includegraphics[width=7cm]{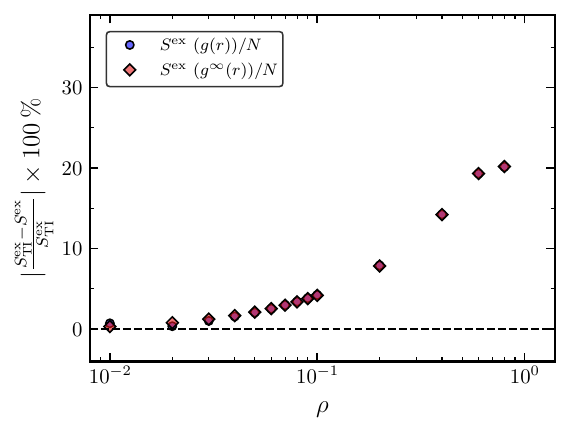}}
    \label{fig7d}
    \caption{Comparison of excess entropies $S^{\mathrm{ex}}$ computed for various densities $\rho$ using Thermodynamic Integration (TI) and Eq.~\ref{eq:excess}, with and without finite-size corrections to the RDF, $g^{\infty}(r)$ and $g(r)$, in the $NVT$ ensemble for $T = 2$, $r_c = 2$, $\alpha = 0.5$, and for different system size $N$: (a) Comparison of $S^{\mathrm{ex}}$ for $N = 100$. (b) Computed Absolute Percentage Error (APE) of $S^{\mathrm{ex}}$ for $N = 100$. (c) Comparison of $S^{\mathrm{ex}}$ for $N = 500$. (d) Computed APE of $S^{\mathrm{ex}}$ for $N = 500$.}
    \label{fig7}
\end{figure}

\begin{figure}[tbh!]
    \captionsetup{justification=raggedright,singlelinecheck=false}
    \centering
     \captionsetup[subfigure]{position=top,labelfont={normal},textfont=normal,singlelinecheck=off,justification=raggedright}
    \subfloat[]{\includegraphics[width=7cm]{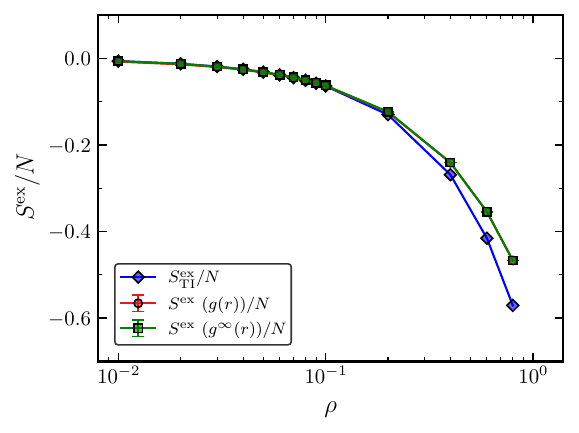}}
    \label{fig8a}
    \subfloat[]{\includegraphics[width=7cm]{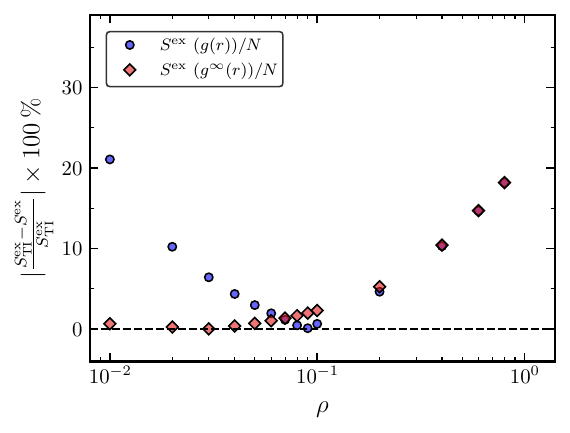}}
    \label{fig8b}
    \subfloat[]{\includegraphics[width=7cm]{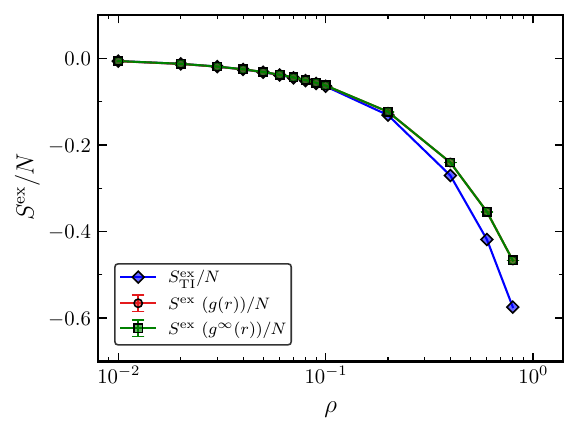}}
    \label{fig8c}
    \subfloat[]{\includegraphics[width=7cm]{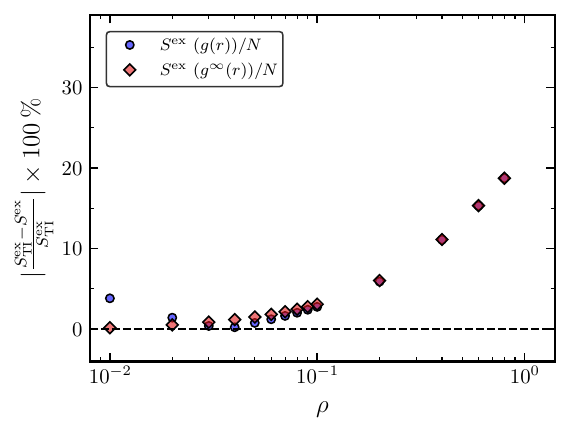}}
    \label{fig8d}
    \caption{Comparison of excess entropies $S^{\mathrm{ex}}$ computed for various densities $\rho$ using Thermodynamic Integration (TI) and Eq.~\ref{eq:excess}, with and without finite-size corrections to the RDF, $g^{\infty}(r)$ and $g(r)$, in the $NVT$ ensemble for $T = 4$, $r_c = 1.2$, $\alpha = 1$, and for different system size $N$: (a) Comparison of $S^{\mathrm{ex}}$ for $N = 100$. (b) Computed Absolute Percentage Error (APE) of $S^{\mathrm{ex}}$ for $N = 100$. (c) Comparison of $S^{\mathrm{ex}}$ for $N = 500$. (d) Computed APE of $S^{\mathrm{ex}}$ for $N = 500$.}
    \label{fig8}
\end{figure}

All MC Simulations were performed for densities $\rho$ ranging from 0.01 to 0.8 in the $NVT$ ensemble. The computed average values of $S^{\mathrm{ex}}$ for different densities, temperatures, and system sizes are shown in Tables~\ref{TAB:1,RC=2;T=4;N=100}-\ref{TAB:6,RC=1.2;T=4;N=500} for $r_c$ $=$ $2$ and $r_c$ $=$ $1.2$. The statistical uncertainties of $S^{\mathrm{ex}}$, computed from all MC simulations are $ < 10^{-3}$. As a result, uncertainties appear smaller than symbols in Fig.~\ref{fig6}a, \ref{fig6}c, \ref{fig7}a, \ref{fig7}c, \ref{fig8}a and \ref{fig8}c. For the sake of clarity, the statistical uncertainties of $S^{\mathrm{ex}}$ is not included in Tables~\ref{TAB:1,RC=2;T=4;N=100}-\ref{TAB:6,RC=1.2;T=4;N=500}. Excess entropies computed from TI ($S^{\mathrm{ex}}_{\rm TI}$) and by integrating RDFs ($S^{\mathrm{ex}}$ using $g(r)$ and $S^{\mathrm{ex}}$ using $g^{\infty}(r)$) for $T=4$ with $r_c=2$, $\alpha=1$ and $N=100$ particles are shown in Table~\ref{TAB:1,RC=2;T=4;N=100}. Simulations were performed for $N =100$ and $N =500$ particles to analyze the effects of system size, and the computed $S^{\mathrm{ex}}$ from TI and RDFs are shown in Table~\ref{TAB:2,RC=2;T=4;N=500}. The computed $S^{\mathrm{ex}}$ listed in Table~\ref{TAB:1,RC=2;T=4;N=100} and~\ref{TAB:2,RC=2;T=4;N=500} are plotted in Fig.~\ref{fig6}a and ~\ref{fig6}c for $N=100$ and $N=500$ particles, respectively. From Fig.~\ref{fig6}a and \ref{fig6}c it is clear that $S^{\mathrm{ex}} \rightarrow 0$ for $\rho \rightarrow 0$. The computed values of $S^{\mathrm{ex}}$ obtained by integrating RDFs ($ g(r)$ and $g^{\infty}(r)$) appear to have excellent agreement at low densities. However, due to the extended axis range in Fig.~\ref{fig6}a and \ref{fig6}c, the discrepancies in $S^{\mathrm{ex}}$ computed by integrating the uncorrected RDF ($g(r)$) are not observed distinctly in Fig.~\ref{fig6}a and \ref{fig6}c. To analyze $S^{\mathrm{ex}}$ computed by integrating RDFs with $S_{\mathrm{TI}}^{\mathrm{ex}}$, Absolute Percentage Errors (APEs) of $S^{\mathrm{ex}}$ computed from RDFs are plotted in Fig.~\ref{fig6}b and \ref{fig6}d for $N=100$ and $N=500$ particles, respectively. These are defined as:
\begin{equation}
    \text{Absolute Percentage Error (APE)} = \left| \frac{S^{\mathrm{ex}}_{\text{TI}} - S^{\mathrm{ex}}}{S^{\mathrm{ex}}_{\text{TI}}} \right| \times 100\%.
    \label{eq:19}
\end{equation}
From Fig.~\ref{fig6}b, it is observed that at low densities, $S^{\mathrm{ex}}$ computed by integrating the RDF corrected to the thermodynamic limit, $g^{\infty}(r)$ provides accurate estimations of $S^{\mathrm{ex}}$. $S^{\mathrm{ex}}$ computed by integrating the RDF without applying the correction for the thermodynamic limit ($g(r)$) has significant absolute percentage errors. These errors become more pronounced when densities decrease in a system with $100$ particles. When the system size is increased from $N=100$ to $N=500$ particles, absolute percentage errors of $S^{\mathrm{ex}}$ obtained using $g(r)$ becomes negligible. Consequently, both $g(r)$ and $g^{\infty}(r)$ provide nearly identical estimations of $S^{\mathrm{ex}}$ for a system consisting of $N=500$ particles. A minor deviation in $S^{\mathrm{ex}}$ (using $g(r)$) is observed at $\rho = 10^{-2}$ in Fig.~\ref{fig6}d, implying that for $\rho < 10^{-2} $, accurate estimation of $S^{\mathrm{ex}}$ from $g(r)$ necessitate system with sizes $ >500$ particles. This illustrates the finite-size effects of $g(r)$, and for very low densities, one needs to consider very large systems for $S^{\mathrm{ex}}$ computation using $g(r)$. Nevertheless, $S^{\mathrm{ex}}$ can be computed accurately with small system sizes (even with $N=100$) by using the RDF corrected to thermodynamic limit ($g^{\infty}(r)$) at low densities. The difference between corrected and uncorrected RDFs computed at $\rho = 10^{-2} $ for $T = 4$, $r_c=2$, and $\alpha=1$ are plotted in Fig.~\ref{fig-RevisionPlot}a for a system with $N=100$ and in Fig.~\ref{fig-RevisionPlot}b for $N=500$ particles. It is clear from Fig.~\ref{fig-RevisionPlot}a that the RDF without finite-size correction differs from the RDF corrected to the thermodynamic limit. The difference lies in the convergence to an asymptotic value of 1, where $g^{\infty}(r)$ reaches 1 at $ r_c = 2 $, while $g(r)$ does not converge precisely to 1. In the case of a system with $N=500$ particles seen in Fig.~\ref{fig-RevisionPlot}b, both RDFs are almost indistinguishable and converge to an asymptotic value of 1 at $r_c=2$. Finite-size effects of $g(r)$ account for the observed differences in the computed $ S^{\mathrm{ex}}$ in Fig.~\ref{fig6}b at low densities. For $\rho \gtrsim 0.1$ absolute percentage errors of $S^{\mathrm{ex}}$ computed from RDFs are $>5\%$ for both $N=100$ and $500$ particles. For $\rho \gtrsim 0.1 $, absolute percentage errors of $S^{\mathrm{ex}}$ computed from RDFs tend to increase as $\rho$ increases. At $\rho$ $=$ $0.8$, absolute percentage errors of $S^{\mathrm{ex}}$ shown in Fig.~\ref{fig6}b and ~\ref{fig6}d are noticed to be $\gtrsim$ $25\%$.

\medskip

\par The values of $S^{\mathrm{ex}}$ computed at $T=2$ with $r_c=2$ and $\alpha=0.5$ for a system with $N=100$ and $N=500$ particles are presented in Table~\ref{TAB:3,RC=2;T=2;N=100} and \ref{TAB:4,RC=2;T=2;N=500}, respectively. For $T=2$, $\alpha$ was chosen as $0.5$ instead of $1$ to avoid a vapor-liquid phase transition during the thermodynamic integration. $S^{\mathrm{ex}}$ computed from TI and by integrating the RDFs are plotted in Fig.~\ref{fig7}a ($N=100$) and Fig.~\ref{fig7}c ($N=500$) including the computed absolute percentage error in Fig.~\ref{fig7}b ($N=100$) and Fig.~\ref{fig7}d ($N=500$). Similar to $T=4$, $S^{\mathrm{ex}}$ computed by integrating the $ g^{\infty}(r)$ are accurate compared to $S^{\mathrm{ex}}_{\rm TI}$ at low densities. $S^{\mathrm{ex}}$ computed by integrating the $g(r)$ for a system with $N=100$ particles suffer from significant absolute percentage errors at low densities, as seen in Fig.~\ref{fig7}b. For $N=500$, both $S^{\mathrm{ex}}$ (using $g^{\infty}(r)$) and $S^{\mathrm{ex}}$ (using $g(r)$) are nearly identical as observed in Fig.~\ref{fig7}d. Absolute percentage errors at $\rho = 10^{-2}$ for $T=2$ and $T=4$ are ca. $6\%$ and ca. $9\%$ for $100$ particles, indicating that APEs of $S^{\mathrm{ex}}$ computed by integrating the $g(r)$ is temperature dependent. Nevertheless, $S^{\mathrm{ex}}$ (using $g^{\infty}(r)$) provides accurate estimation independent of temperature. Absolute percentage errors of $S^{\mathrm{ex}}$ (using $g^{\infty}(r)$ and $g(r)$) at high densities are found to be increasing with increasing density (maximum of $\approx20\%$ at $\rho$ = 0.8). MC simulations were also performed to compute $S^{\mathrm{ex}}$ for $r_c =1.2$, where the WF potential behaves like colloid particles. The computed $S^{\mathrm{ex}}$ from TI and RDFs for $T=4$ with $N=100$ particles are listed and plotted in Table~\ref{TAB:5,RC=1.2;T=4;N=100} and Fig.~\ref{fig8}a, respectively. $S^{\mathrm{ex}}$ computed by integrating $g^{\infty}(r)$ were observed to be in agreement with $S^{\mathrm{ex}}_{\rm TI}$ at low densities, with discrepancies reaching up to $20\%$ at high densities. Absolute percentage error of $S^{\mathrm{ex}}$ (using $g(r)$) at $\rho=10^{-2}$ in Fig.~\ref{fig8}b is $\gtrsim 21\% $, whereas in Fig.~\ref{fig6}b ($r_c =2$ and $N=100$) the APE was $\gtrsim 9\% $. This indicates that $S^{\mathrm{ex}}$ (using $g(r)$) suffers substantial inaccuracies for colloid-like particles ($r_c =1.2$) compared to Lennard-Jones-like particles ($r_c =2$) at low densities. For a system size with $N=500$ particles the computed $S^{\mathrm{ex}}$ are plotted in Fig.~\ref{fig8}c (also listed in Table~\ref{TAB:6,RC=1.2;T=4;N=500}) and their corresponding absolute percentage errors in Fig.~\ref{fig8}d. For extremely low densities ($\rho$ $=$ $10^{-2}$), even for a system of $500$ particles $S^{\mathrm{ex}}$ (using $g(r)$) show notable errors ($\approx$ 4\%) compared to $S^{\mathrm{ex}}_{\rm TI}$ and $S^{\mathrm{ex}}$ (using $g^{\infty}(r)$) in Fig.~\ref{fig8}d. It is clear from this that the corrected RDF $g^{\infty}(r)$ should be used in the $S^{\mathrm{ex}}$ computation for both $r_c = 2$ and $r_c = 1.2$ to obtain accurate $S^{\mathrm{ex}}_{\rm TI}$ with small system sizes at low densities. Comparing the values of $S^{\mathrm{ex}}$ computed using TI for systems with $N =100$ and $500$ particles show that $S^{\mathrm{ex}}_{\rm TI}$ is independent of system size regardless of $r_c$. The maximum difference of $S^{\mathrm{ex}}_{\rm TI}$ between different system sizes was found to be $\approx$ $1.6\%$ for $T$ $=$ $4$, $\rho$ $=$ $0.8$, and $r_c$ $=$ $2$. The discrepancies of $S^{\mathrm{ex}}$ computed using RDFs observed at high densities in Figs.~\ref{fig6}-\ref{fig8} can be attributed to the low-density approximation inherent in the second-order density expansion of entropy (Eq.~\ref{eq:excess}). The higher-order density expansion of entropy can be used to compute $S^{\mathrm{ex}}$ at high densities by following the approximations proposed by Huang and Widom~\cite{huang2024entropy}. However, these approximations lose validity in the vicinity of the liquid-to-solid transition ($\rho\gtrsim0.8$)~\cite{huang2024entropy}. Including higher-order terms could lead to high discrepancies compared to second-order near the liquid-to-solid transition~\cite{huang2024entropy}.

\section{Conclusions}
\label{sec:conclusions}

In this paper, we have investigated the computation of $S^{\mathrm{ex}}$ and KB integrals extrapolated to the thermodynamic limit ($L \rightarrow \infty $) in a finite volume using analytic RDFs. Expressions to compute $S^{\mathrm{ex}}$ and KB integral at $L \rightarrow \infty$ were derived based on a Taylor expansion in $1/L$ for different orders. We observed that the running integral $X^*(L)$ (Eq.~\ref{eq:5}) and first-order approximation $X_\infty^1$ (Eq.~\ref{eq:10}) converge faster than other approximations for $S^{\mathrm{ex}}$ and KB integrals for different ranges of $g(r)$. We noticed that the $X^*(L)$ approximation integral for $S^{\mathrm{ex}}$ converged much faster than the $X_\infty^1$ approximation of the KB integral, irrespective of the range of $g(r)$. We showed that truncation of $S^{\mathrm{ex}}$ and KB integrals is possible, provided the appropriate approximated expression and $L$ are chosen based on the range of $g(r)$. We also investigated finite-size effects of the RDF in computing $S^{\mathrm{ex}}$ from MC simulations using the WF potential. We found that $S^{\mathrm{ex}}$ computed by integrating RDF corrected to the thermodynamic limit ($g^{\infty}(r)$) agrees with $S^{\mathrm{ex}}$ computed from thermodynamic integration ($S^{\mathrm{ex}}_{\rm TI}$) for both Lennard-Jones-like ($r_c =2$) and colloid-like ($r_c =1.2$) particles at low densities. This agreement holds for systems with $100$ and $500$ particles. We noticed that $S_{\rm TI}^{\mathrm{ex}}$ computed by thermodynamic integration showed no significant difference in the values of $S^{\mathrm{ex}}$ for different system sizes. $S^{\mathrm{ex}}$ computed by integrating the standard RDF ($g(r)$) show significant discrepancies at low densities for a system with $100$ particles (for both $r_c =2$ and $r_c =1.2$). For a system size of $500$ particles, $S^{\mathrm{ex}}$ computed by integrating $g(r)$ showed minor discrepancies at extremely low densities for both $r_c =2$ and $r_c =1.2$ suggesting that $g^{\infty}(r)$ should always be used in the computation of $S^{\mathrm{ex}}$. At high densities ($\rho > 0.1$), $S^{\mathrm{ex}}$ computed by integrating RDFs ($g^{\infty}(r)$ and $g(r)$) yields identical values. Comparing $S^{\mathrm{ex}}_{\rm TI}$ with $S^{\mathrm{ex}}$ computed from RDFs shows significant differences for $\rho > 0.1$. Discrepancies at high densities are due to the second-order approximation of the excess entropy integral (Eq.~\ref{eq:excess}) used for computing $S^{\mathrm{ex}}$ from MC simulations. The computation of $S^{\mathrm{ex}}$ using Eq.~\ref{eq:excess} truncated to a finite-size and $g^{\infty}(r)$ captures $95\%$ of the $S^{\mathrm{ex}}$ for $\rho< 0.1$. This level of accuracy in $S^{\mathrm{ex}}$ computation holds for both Lennard-Jones-like and colloid-like particles for $\rho < 0.1$, even for a small system size of $100$ particles. Our simulation results indicate that accurate estimations of $S^{\mathrm{ex}}$ can be obtained from Eq.~\ref{eq:excess} and TI for $\rho < 0.1$ for a system with $100$ particles. The computation of $S^{\mathrm{ex}}$ using Eq.~\ref{eq:excess} and $g^{\infty}(r)$ can result in errors of $20\%$ at high densities, regardless of $r_c$. A summary of the comparison and observations related to the excess entropy computation investigated in this study is presented in Table~\ref{TAB:7}. Our approach enables the efficient and computationally inexpensive computation of $S^{\mathrm{ex}}$ by addressing the underlying approximations. 

\newpage
\raggedright{\textbf{Acknowledgments}} \\
The work presented herein is part of the ENCASE project (A European Network of Research Infrastructures for \ce{CO2} Transport and Injection). ENCASE has received funding from the European Union’s Horizon Europe Research and Innovation program under grant Number 101094664. This work was also sponsored by NWO domain Science for the use of supercomputer facilities, with financial support from the Nederlandse Organisatie voor Wetenschappelijk Onderzoek (The Netherlands Organization for Scientific Research, NWO). The authors acknowledge the use of computational resources of the DelftBlue supercomputer, provided by Delft High Performance Computing Center (https://www.tudelft.nl/dhpc).

\newpage
\clearpage
\bibliographystyle{tfq}
\bibliography{paper}

\newpage

\renewcommand{\arraystretch}{1.5} 
\newcolumntype{P}[1]{>{\centering\arraybackslash}m{#1}}

\begin{table}[tbh!]
\caption{Excess entropies $S^{\mathrm{ex}}$ computed for various densities $\rho$ from Thermodynamic Integration (TI) using $A^{\rm ex} = U-T S^{\rm ex}$ and Eq.~\ref{eq:excess}, with and without finite-size corrections to the Radial Distribution Function (RDF). The corrected RDF $g^{\infty}(r)$ (Eq.~\ref{eq:18}) uses the method proposed by Ganguly and van der Vegt~\cite{ganguly2013convergence}, while the uncorrected one uses the RDF $g(r)$ directly. Simulations were performed in the $NVT$ ensemble for $T = 4$, $r_c = 2$, $\alpha = 1$, and $N = 100$.}
\label{TAB:1,RC=2;T=4;N=100}
\centering
\begin{tabular}{P{0.7cm} P{1.75cm} P{1.75cm} P{1.75cm} P{1.75cm} P{1.75cm} P{1.75cm}} 
\toprule
${ \rho }$ & {$ U/N $} & {$ S^{\mathrm{ex}}_{\mathrm{TI}}/N $} & \shortstack{$ S^{\mathrm{ex}}$ \\ $(g^{\infty}(r))/N $} & \shortstack{$ S^{\mathrm{ex}}$ \\ $(g(r))/N $} &  \shortstack{ Absolute \\  Percentage \\ Error - \\ $S^{\mathrm{ex}}~(g^{\infty}(r))$} & \shortstack{Absolute \\  Percentage \\ Error -  \\ $S^{\mathrm{ex}}~(g(r))$} \\ \midrule 
 0.01                         & -0.0391                         & -0.0144                         & -0.0145                         & -0.0157                         & 0.54                         & 8.81 \\ 
 0.02                         & -0.0782                         & -0.0290                         & -0.0290                         & -0.0301                         & 0.07                         & 3.76 \\ 
 0.03                         & -0.1170                         & -0.0436                         & -0.0434                         & -0.0444                         & 0.36                         & 1.84 \\ 
 0.04                         & -0.1558                         & -0.0583                         & -0.0578                         & -0.0587                         & 0.80                         & 0.67 \\ 
 0.05                         & -0.1944                         & -0.0731                         & -0.0722                         & -0.0730                         & 1.23                         & 0.19 \\ 
 0.06                         & -0.2328                         & -0.0880                         & -0.0865                         & -0.0872                         & 1.67                         & 0.91 \\ 
 0.07                         & -0.2712                         & -0.1030                         & -0.1009                         & -0.1014                         & 2.09                         & 1.53 \\ 
 0.08                         & -0.3093                         & -0.1181                         & -0.1151                         & -0.1156                         & 2.51                         & 2.10 \\ 
 0.09                         & -0.3473                         & -0.1333                         & -0.1294                         & -0.1298                         & 2.94                         & 2.63 \\ 
 0.10                         & -0.3852                         & -0.1486                         & -0.1436                         & -0.1440                         & 3.37                         & 3.14 \\ 
 0.20                         & -0.7561                         & -0.3083                         & -0.2852                         & -0.2850                         & 7.49                         & 7.55 \\ 
 0.40                         & -1.4341                         & -0.6701                         & -0.5694                         & -0.5691                         & 15.02                         & 15.07 \\ 
 0.60                         & -1.9117                         & -1.0982                         & -0.8707                         & -0.8712                         & 20.71                         & 20.67 \\ 
 0.80                         & -1.8920                         & -1.5865                         & -1.2153                         & -1.2168                         & 23.40                         & 23.30 \\ 
\bottomrule 
\end{tabular} 
\end{table}

\begin{table}[tbh!]
\caption{Excess entropies $S^{\mathrm{ex}}$ computed for various densities $\rho$ from Thermodynamic Integration (TI) using $A^{\rm ex} = U-T S^{\rm ex}$ and Eq.~\ref{eq:excess}, with and without finite-size corrections to the Radial Distribution Function (RDF). The corrected RDF $g^{\infty}(r)$ (Eq.~\ref{eq:18}) uses the method proposed by Ganguly and van der Vegt~\cite{ganguly2013convergence}, while the uncorrected one uses the RDF $g(r)$ directly. Simulations were performed in the $NVT$ ensemble for $T = 4$, $r_c = 2$, $\alpha = 1$, and $N = 500$.}
\label{TAB:2,RC=2;T=4;N=500}
\centering
\begin{tabular}{P{0.7cm} P{1.75cm} P{1.75cm} P{1.75cm} P{1.75cm} P{1.75cm} P{1.75cm}} 
\toprule
${ \rho }$ & {$ U/N $} & {$ S^{\mathrm{ex}}_{\mathrm{TI}}/N $} & \shortstack{$ S^{\mathrm{ex}}$ \\ $(g^{\infty}(r))/N $} & \shortstack{$ S^{\mathrm{ex}}$ \\ $(g(r))/N $} &  \shortstack{ Absolute \\  Percentage \\ Error - \\ $S^{\mathrm{ex}}~(g^{\infty}(r))$} & \shortstack{Absolute \\  Percentage \\ Error -  \\ $S^{\mathrm{ex}}~(g(r))$} \\ \midrule 
 0.01                         & -0.0395                         & -0.0146                         & -0.0145                         & -0.0147                         & 0.28                         & 1.35 \\ 
 0.02                         & -0.0788                         & -0.0292                         & -0.0290                         & -0.0292                         & 0.73                         & 0.00 \\ 
 0.03                         & -0.1179                         & -0.0439                         & -0.0434                         & -0.0436                         & 1.15                         & 0.71 \\ 
 0.04                         & -0.1570                         & -0.0588                         & -0.0578                         & -0.0580                         & 1.59                         & 1.29 \\ 
 0.05                         & -0.1958                         & -0.0737                         & -0.0722                         & -0.0724                         & 2.01                         & 1.80 \\ 
 0.06                         & -0.2345                         & -0.0887                         & -0.0865                         & -0.0867                         & 2.43                         & 2.28 \\ 
 0.07                         & -0.2731                         & -0.1038                         & -0.1009                         & -0.1010                         & 2.85                         & 2.74 \\ 
 0.08                         & -0.3115                         & -0.1190                         & -0.1152                         & -0.1153                         & 3.27                         & 3.18 \\ 
 0.09                         & -0.3498                         & -0.1344                         & -0.1294                         & -0.1295                         & 3.68                         & 3.61 \\ 
 0.10                         & -0.3879                         & -0.1498                         & -0.1437                         & -0.1437                         & 4.09                         & 4.05 \\ 
 0.20                         & -0.7601                         & -0.3106                         & -0.2853                         & -0.2853                         & 8.13                         & 8.14 \\ 
 0.40                         & -1.4370                         & -0.6744                         & -0.5699                         & -0.5698                         & 15.49                         & 15.50 \\ 
 0.60                         & -1.9096                         & -1.1042                         & -0.8722                         & -0.8723                         & 21.01                         & 21.00 \\ 
 0.80                         & -1.8811                         & -1.6118                         & -1.2188                         & -1.2192                         & 24.38                         & 24.36 \\ 
\bottomrule 
\end{tabular} 
\end{table}

\begin{table}[tbh!]
\caption{Excess entropies $S^{\mathrm{ex}}$ computed for various densities $\rho$ from Thermodynamic Integration (TI) using $A^{\rm ex} = U-T S^{\rm ex}$ and Eq.~\ref{eq:excess}, with and without finite-size corrections to the Radial Distribution Function (RDF). The corrected RDF $g^{\infty}(r)$ (Eq.~\ref{eq:18}) uses the method proposed by Ganguly and van der Vegt~\cite{ganguly2013convergence}, while the uncorrected one uses the RDF $g(r)$ directly. Simulations were performed in the $NVT$ ensemble for $T = 2$, $r_c = 2$, $\alpha = 0.5$, and $N = 100$.}
\label{TAB:3,RC=2;T=2;N=100}
\centering
\begin{tabular}{P{0.7cm} P{1.75cm} P{1.75cm} P{1.75cm} P{1.75cm} P{1.75cm} P{1.75cm}} 
\toprule
${ \rho }$ & {$ U/N $} & {$ S^{\mathrm{ex}}_{\mathrm{TI}}/N $} & \shortstack{$ S^{\mathrm{ex}}$ \\ $(g^{\infty}(r))/N $} & \shortstack{$ S^{\mathrm{ex}}$ \\ $(g(r))/N $} &  \shortstack{ Absolute \\  Percentage \\ Error - \\ $S^{\mathrm{ex}}~(g^{\infty}(r))$} & \shortstack{Absolute \\  Percentage \\ Error -  \\ $S^{\mathrm{ex}}~(g(r))$} \\ \midrule 
 0.01                         & -0.0538                         & -0.0198                         & -0.0199                         & -0.0210                         & 0.51                         & 5.97 \\ 
 0.02                         & -0.1072                         & -0.0396                         & -0.0396                         & -0.0404                         & 0.00                         & 2.17 \\ 
 0.03                         & -0.1602                         & -0.0593                         & -0.0591                         & -0.0597                         & 0.47                         & 0.62 \\ 
 0.04                         & -0.2129                         & -0.0791                         & -0.0784                         & -0.0789                         & 0.92                         & 0.36 \\ 
 0.05                         & -0.2652                         & -0.0989                         & -0.0976                         & -0.0978                         & 1.37                         & 1.11 \\ 
 0.06                         & -0.3172                         & -0.1188                         & -0.1166                         & -0.1167                         & 1.81                         & 1.75 \\ 
 0.07                         & -0.3687                         & -0.1386                         & -0.1355                         & -0.1354                         & 2.23                         & 2.30 \\ 
 0.08                         & -0.4200                         & -0.1584                         & -0.1542                         & -0.1539                         & 2.65                         & 2.82 \\ 
 0.09                         & -0.4709                         & -0.1782                         & -0.1728                         & -0.1724                         & 3.07                         & 3.30 \\ 
 0.10                         & -0.5214                         & -0.1981                         & -0.1912                         & -0.1907                         & 3.48                         & 3.75 \\ 
 0.20                         & -1.0116                         & -0.4002                         & -0.3713                         & -0.3700                         & 7.22                         & 7.56 \\ 
 0.40                         & -1.9391                         & -0.8484                         & -0.7311                         & -0.7306                         & 13.83                         & 13.89 \\ 
 0.60                         & -2.7991                         & -1.4100                         & -1.1385                         & -1.1397                         & 19.26                         & 19.17 \\ 
 0.80                         & -3.3304                         & -2.0935                         & -1.6624                         & -1.6644                         & 20.59                         & 20.50 \\ 
\bottomrule 
\end{tabular} 
\end{table}

\begin{table}[tbh!]
\caption{Excess entropies $S^{\mathrm{ex}}$ computed for various densities $\rho$ from Thermodynamic Integration (TI) using $A^{\rm ex} = U-T S^{\rm ex}$ and Eq.~\ref{eq:excess}, with and without finite-size corrections to the Radial Distribution Function (RDF). The corrected RDF $g^{\infty}(r)$ (Eq.~\ref{eq:18}) uses the method proposed by Ganguly and van der Vegt~\cite{ganguly2013convergence}, while the uncorrected one uses the RDF $g(r)$ directly. Simulations were performed in the $NVT$ ensemble for $T = 2$, $r_c = 2$, $\alpha = 0.5$, and $N = 500$.}
\label{TAB:4,RC=2;T=2;N=500}
\centering
\begin{tabular}{P{0.7cm} P{1.75cm} P{1.75cm} P{1.75cm} P{1.75cm} P{1.75cm} P{1.75cm}} 
\toprule
${ \rho }$ & {$ U/N $} & {$ S^{\mathrm{ex}}_{\mathrm{TI}}/N $} & \shortstack{$ S^{\mathrm{ex}}$ \\ $(g^{\infty}(r))/N $} & \shortstack{$ S^{\mathrm{ex}}$ \\ $(g(r))/N $} &  \shortstack{ Absolute \\  Percentage \\ Error - \\ $S^{\mathrm{ex}}~(g^{\infty}(r))$} & \shortstack{Absolute \\  Percentage \\ Error -  \\ $S^{\mathrm{ex}}~(g(r))$} \\ \midrule 
 0.01                         & -0.0542                         & -0.0199                         & -0.0199                         & -0.0201                         & 0.34                         & 0.74 \\ 
 0.02                         & -0.1081                         & -0.0399                         & -0.0396                         & -0.0398                         & 0.81                         & 0.38 \\ 
 0.03                         & -0.1616                         & -0.0599                         & -0.0591                         & -0.0592                         & 1.26                         & 1.04 \\ 
 0.04                         & -0.2146                         & -0.0798                         & -0.0785                         & -0.0785                         & 1.70                         & 1.59 \\ 
 0.05                         & -0.2674                         & -0.0998                         & -0.0977                         & -0.0977                         & 2.13                         & 2.08 \\ 
 0.06                         & -0.3198                         & -0.1198                         & -0.1167                         & -0.1167                         & 2.57                         & 2.55 \\ 
 0.07                         & -0.3718                         & -0.1398                         & -0.1356                         & -0.1356                         & 2.99                         & 3.00 \\ 
 0.08                         & -0.4234                         & -0.1598                         & -0.1544                         & -0.1543                         & 3.40                         & 3.43 \\ 
 0.09                         & -0.4748                         & -0.1799                         & -0.1730                         & -0.1729                         & 3.80                         & 3.85 \\ 
 0.10                         & -0.5258                         & -0.1999                         & -0.1915                         & -0.1914                         & 4.20                         & 4.25 \\ 
 0.20                         & -1.0193                         & -0.4039                         & -0.3722                         & -0.3719                         & 7.86                         & 7.92 \\ 
 0.40                         & -1.9470                         & -0.8547                         & -0.7330                         & -0.7328                         & 14.24                         & 14.25 \\ 
 0.60                         & -2.8001                         & -1.4171                         & -1.1431                         & -1.1434                         & 19.33                         & 19.32 \\ 
 0.80                         & -3.3181                         & -2.0996                         & -1.6758                         & -1.6763                         & 20.19                         & 20.16 \\ 
\bottomrule 
\end{tabular} 
\end{table}

\begin{table}[tbh!]
\caption{Excess entropies $S^{\mathrm{ex}}$ computed for various densities $\rho$ from Thermodynamic Integration (TI) using $A^{\rm ex} = U-T S^{\rm ex}$ and Eq.~\ref{eq:excess}, with and without finite-size corrections to the Radial Distribution Function (RDF). The corrected RDF $g^{\infty}(r)$ (Eq.~\ref{eq:18}) uses the method proposed by Ganguly and van der Vegt~\cite{ganguly2013convergence}, while the uncorrected one uses the RDF $g(r)$ directly. Simulations were performed in the $NVT$ ensemble for $T = 4$, $r_c = 1.2$, $\alpha = 1$, and $N = 100$.}
\label{TAB:5,RC=1.2;T=4;N=100}
\centering
\begin{tabular}{P{0.7cm} P{1.75cm} P{1.75cm} P{1.75cm} P{1.75cm} P{1.75cm} P{1.75cm}} 
\toprule
${ \rho }$ & {$ U/N $} & {$ S^{\mathrm{ex}}_{\mathrm{TI}}/N $} & \shortstack{$ S^{\mathrm{ex}}$ \\ $(g^{\infty}(r))/N $} & \shortstack{$ S^{\mathrm{ex}}$ \\ $(g(r))/N $} &  \shortstack{ Absolute \\  Percentage \\ Error - \\ $S^{\mathrm{ex}}~(g^{\infty}(r))$} & \shortstack{Absolute \\  Percentage \\ Error -  \\ $S^{\mathrm{ex}}~(g(r))$} \\ \midrule 
 0.01                         & 0.0087                         & -0.0063                         & -0.0064                         & -0.0077                         & 0.68                         & 21.08 \\ 
 0.02                         & 0.0175                         & -0.0127                         & -0.0127                         & -0.0140                         & 0.28                         & 10.24 \\ 
 0.03                         & 0.0264                         & -0.0190                         & -0.0190                         & -0.0202                         & 0.05                         & 6.45 \\ 
 0.04                         & 0.0353                         & -0.0254                         & -0.0253                         & -0.0265                         & 0.40                         & 4.38 \\ 
 0.05                         & 0.0444                         & -0.0318                         & -0.0316                         & -0.0328                         & 0.73                         & 3.00 \\ 
 0.06                         & 0.0536                         & -0.0382                         & -0.0378                         & -0.0390                         & 1.06                         & 1.98 \\ 
 0.07                         & 0.0628                         & -0.0447                         & -0.0441                         & -0.0452                         & 1.38                         & 1.18 \\ 
 0.08                         & 0.0722                         & -0.0511                         & -0.0503                         & -0.0514                         & 1.69                         & 0.49 \\ 
 0.09                         & 0.0817                         & -0.0576                         & -0.0565                         & -0.0575                         & 2.01                         & 0.13 \\ 
 0.10                         & 0.0912                         & -0.0641                         & -0.0626                         & -0.0637                         & 2.33                         & 0.67 \\ 
 0.20                         & 0.1928                         & -0.1303                         & -0.1234                         & -0.1242                         & 5.28                         & 4.66 \\ 
 0.40                         & 0.4312                         & -0.2687                         & -0.2406                         & -0.2410                         & 10.45                         & 10.32 \\ 
 0.60                         & 0.7257                         & -0.4155                         & -0.3543                         & -0.3544                         & 14.72                         & 14.71 \\ 
 0.80                         & 1.0882                         & -0.5705                         & -0.4667                         & -0.4666                         & 18.20                         & 18.21 \\ 
\bottomrule 
\end{tabular} 
\end{table}

\begin{table}[tbh!]
\caption{Excess entropies $S^{\mathrm{ex}}$ computed for various densities $\rho$ from Thermodynamic Integration (TI) using $A^{\rm ex} = U-T S^{\rm ex}$ and Eq.~\ref{eq:excess}, with and without finite-size corrections to the Radial Distribution Function (RDF). The corrected RDF $g^{\infty}(r)$ (Eq.~\ref{eq:18}) uses the method proposed by Ganguly and van der Vegt~\cite{ganguly2013convergence}, while the uncorrected one uses the RDF $g(r)$ directly. Simulations were performed in the $NVT$ ensemble for $T = 4$, $r_c = 1.2$, $\alpha = 1$, and $N = 500$.}
\label{TAB:6,RC=1.2;T=4;N=500}
\centering
\begin{tabular}{P{0.7cm} P{1.75cm} P{1.75cm} P{1.75cm} P{1.75cm} P{1.75cm} P{1.75cm}} 
\toprule
${ \rho }$ & {$ U/N $} & {$ S^{\mathrm{ex}}_{\mathrm{TI}}/N $} & \shortstack{$ S^{\mathrm{ex}}$ \\ $(g^{\infty}(r))/N $} & \shortstack{$ S^{\mathrm{ex}}$ \\ $(g(r))/N $} &  \shortstack{ Absolute \\  Percentage \\ Error - \\ $S^{\mathrm{ex}}~(g^{\infty}(r))$} & \shortstack{Absolute \\  Percentage \\ Error -  \\ $S^{\mathrm{ex}}~(g(r))$} \\ \midrule  
 0.01                         & 0.0088                         & -0.0064                         & -0.0064                         & -0.0066                         & 0.19                         & 3.84 \\ 
 0.02                         & 0.0176                         & -0.0128                         & -0.0127                         & -0.0129                         & 0.54                         & 1.44 \\ 
 0.03                         & 0.0266                         & -0.0192                         & -0.0190                         & -0.0193                         & 0.88                         & 0.41 \\ 
 0.04                         & 0.0356                         & -0.0256                         & -0.0253                         & -0.0255                         & 1.20                         & 0.25 \\ 
 0.05                         & 0.0447                         & -0.0321                         & -0.0316                         & -0.0318                         & 1.52                         & 0.79 \\ 
 0.06                         & 0.0540                         & -0.0385                         & -0.0378                         & -0.0380                         & 1.85                         & 1.24 \\ 
 0.07                         & 0.0633                         & -0.0450                         & -0.0440                         & -0.0443                         & 2.17                         & 1.66 \\ 
 0.08                         & 0.0728                         & -0.0515                         & -0.0503                         & -0.0505                         & 2.48                         & 2.05 \\ 
 0.09                         & 0.0823                         & -0.0581                         & -0.0564                         & -0.0567                         & 2.80                         & 2.42 \\ 
 0.10                         & 0.0919                         & -0.0646                         & -0.0626                         & -0.0628                         & 3.11                         & 2.78 \\ 
 0.20                         & 0.1941                         & -0.1313                         & -0.1233                         & -0.1235                         & 6.04                         & 5.92 \\ 
 0.40                         & 0.4339                         & -0.2707                         & -0.2405                         & -0.2406                         & 11.14                         & 11.12 \\ 
 0.60                         & 0.7296                         & -0.4184                         & -0.3542                         & -0.3543                         & 15.34                         & 15.33 \\ 
 0.80                         & 1.0932                         & -0.5744                         & -0.4667                         & -0.4667                         & 18.75                         & 18.75 \\ 
\bottomrule 
\end{tabular} 
\end{table}


\begin{table}[tbh!]
\caption{Summary of underlying key assumptions to consider for excess entropy  ($S^{\rm{ex}}$) computation using molecular simulations.}
\label{TAB:7}
\centering
\begin{tabular}{P{1.2cm} >{\raggedright\arraybackslash}p{12.7cm}}
    \toprule
    \textbf{S. No.} & \textbf{Observations and remarks} \\
    \midrule
    1.& In contrast to KB integrals, truncation of excess entropy integrals (Eq.~\ref{eq:excess}) provides better convergence compared to other approximations.\\
    2.& Finite-size corrected RDFs suggested by Ganguly and van der Vegt~\cite{ganguly2013convergence} must be used for computing the excess entropy, irrespective of the system size. \\
    3. & Eq.~\ref{eq:excess} is a low-density approximation, for $\rho$ $>0.1$ thermodynamic integration or other approximation methods suggested by Huang and Widom~\cite{huang2024entropy} using higher-order density expansion of entropy are preferred.\\
    \bottomrule
\end{tabular}
\end{table}

\clearpage
\newpage


\end{document}